\documentclass[twocolumn,showpacs,preprintnumbers,amsmath,amssymb,superscriptaddress]{revtex4}

\usepackage{color}

\usepackage{graphicx}
\usepackage{graphics}

\def\3{2.8in}    %used for figure widths
\def\2{2.5in}
\def\4{3.0in}

\def \beq {\begin{equation}}
\def \eeq {\end{equation}}
\begin{document}

\title{Dirac cone in iron-based superconductors}
\author{M. Zahid Hasan}\affiliation {Joseph Henry Laboratories, Department of Physics, Princeton University, Princeton, New Jersey 08544, USA}
\author{B. Andrei Bernevig}\affiliation {Joseph Henry Laboratories, Department of Physics, Princeton University, Princeton, New Jersey 08544, USA}

%\pacs{}
%76.20.+q,
%85.35.Gv,
%82.20.Xr,
%76.60.Jx,
%39.20.+q}

\begin{abstract}
The band structure of iron-based superconductors gives rise to yet another scenario for the appearance of Dirac fermions. A viewpoint on \textquotedblleft Observation of Dirac cone electronic dispersion in BaFe$_2$As$_2$\textquotedblright: 
(Richard et.al., PRL 104, 137001 (2010)).

%\textcolor{red}

\end{abstract}
\date{\today}

\maketitle

Superconductivity above 30 K was realized in the mid 80s with the discovery of cuprate high-temperature superconductors \cite{Bednorz}. The monopoly of copper was broken in 2008 with the discovery of iron-based superconductors - of which there are now many families - which have a maximum transition temperature (55 K) comparable to that of single-layer cuprates \cite{Kamihara, Ren}. Their complex phase diagrams are poorly understood, but we do know that most have a magnetic phase \cite{Cruz} that changes to a superconducting one as the concentration of electrons or holes in the bulk increases.
%\vspace{0.3cm}

The parent magnetic state of iron-based superconductors determines their superconducting transition temperature. In their paper in \textit{Physical Review Letters} \cite{Richard}, Pierre Richard and collaborators from China, Japan, and the US report band-structure measurements of the parent magnetic state of iron-based superconductors that show linearly dispersed bands in the shape of a cone, which implies that the carriers follow dispersion relations for massless relativistic fermions that obey the Dirac equation. These results confirm an existing theoretical prediction \cite{Ran} and earlier experimental results on the Dirac band structure and the orbital nature of the magnetic order in the parent superconductor \cite{Hsieh}. Richard {\it et al.} present angle-resolved photoemission (ARPES) studies of high-quality single crystals of BaFe$_2$As$_2$, which has a magnetic transition temperature of 138 K when undoped. They map the Fermi surfaces and find slightly anisotropic band crossings arranged in the shape of a Dirac cone. Dirac fermions are also known to exist in other condensed-matter systems such as in the bulk \cite{Hsieh1} and surface \cite{Xia} of topological insulators \cite{hasan, Bernevig}, in graphene \cite{Geim}, and cuprate superconductors \cite{Damascelli}.

\begin{figure*}
%\centering
\includegraphics[width=10cm]{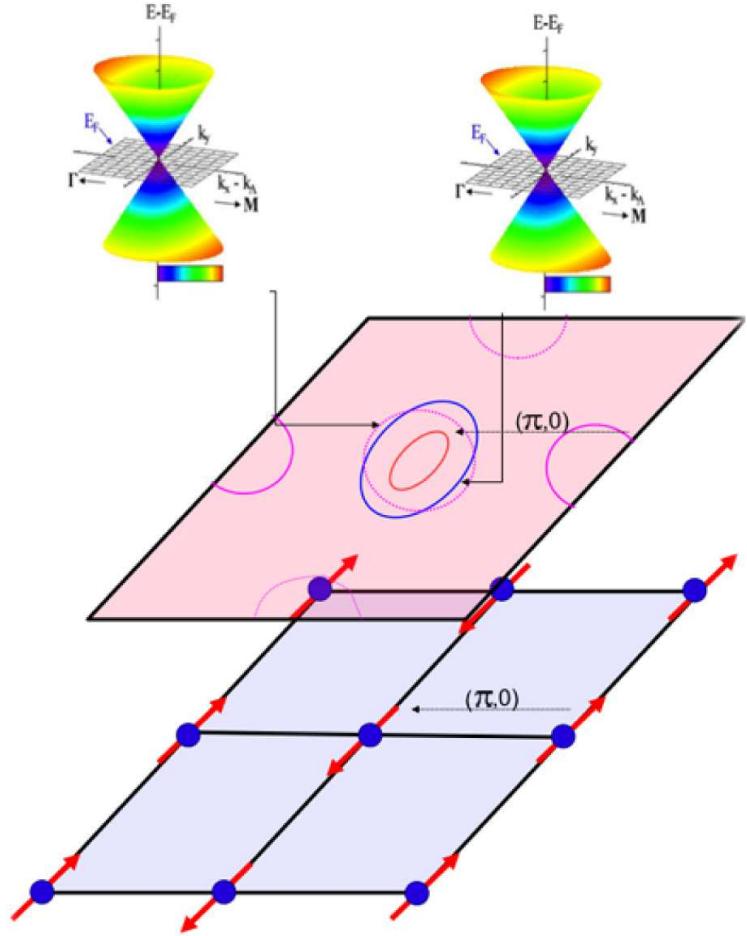}
\caption{The magnetic order and Dirac cones in the pnictides: Spins are projected on a real-space lattice where the magnetic ordering wave vector ($\pi$,0) is shown (bottom plane). This wave vector connects two pieces of the Fermi surface on the reciprocal space (upper plane). The crystal symmetry, multiorbital nature along with the magnetic order in pnictides leaves behind linear band crossings in the shape of Dirac cones in some parts of the reciprocal space plane (upper plane). These Dirac cones highlight the mechanism of magnetic order.}
\end{figure*}

In iron-arsenide compounds, the parent state is a collinear antiferromagnet, with a commensurate ordering wave vector ($\pi$, 0) or (0, $\pi$). This wave vector also appears in the band structure of the normal state of iron-based superconductors as a momentum-space vector (nesting vector) connecting the Fermi surfaceÕs hole pockets that surround the $\Gamma$ point and the electron pockets around the $M$ point in the unfolded Brillouin zone (see Fig. 1). This band structure is a consequence of the multiorbital nature of the system; all of ironÕs five 3$d$ orbitals are necessary for a full electronic description of the system. Of these, the most important for understanding the present experimental results are the \textit{d$_{xz}$, d$_{xy}$, and d$_{yz}$} orbitals.

Plausible explanations for the low-temperature magnetic ordering vector involve either a spin-density wave (SDW) formation tendency consistent with the nesting wave vector \cite{Mazin} or an antiferromagnetic state formation tendency due to next nearest neighbor spin-spin interactions \cite{Fang, Si}. In the magnetic phase, the Fermi surfaces reorganize, the Brillouin zone folds, and gaps due to electron-hole scattering are seen in the electronic spectrum. SDW formation usually leads to the opening of an electronic gap by destroying the Fermi surface if the Bloch vectors along the nesting vectors can span the full Fermi surface. Strong interactions can also gap the SDW state. In the iron-based compounds, as the present experiment shows, the situation is different: due to the topology of the Fermi surface \cite{Singh} a coupling between the Bloch wave vectors of the electron and hole pockets sets the SDW-induced electronic gap.

In accounting for the nature of the SDW state, a full theoretical analysis [6] based on a five-orbital model \cite{Kuroki, Graser} shows that the Bloch band of the electron Fermi surface pocket is odd under reflection, while the outer hole pocket of the Fermi surface is even along the SDW ordering axis. The SDW matrix element between them vanishes and gives rise to a nodal (not fully gapped) SDW, which can at most be semimetallic. This gapless SDW magnetic state survives even for strong interactions \cite{Ran}. In experiments, the authors observe \cite{Richard} a high-intensity point in the ARPES data on the $\Gamma$?M symmetry line for the folded band structure of magnetically ordered BaFe$_2$As$_2$. This feature, occurring roughly where the electron band hybridizes with a hole band, disappears above the SDW transition temperature and has a dispersion characteristic of a relativistic Dirac cone. The observed Dirac cone is thus related to the nature of the SDW state in these materials. The apex of the Dirac cone is situated slightly above the Fermi energy, which would give rise to small hole pockets in the ARPES data. Its dispersion has a small degree of anisotropy. Away from the cone the band is gapped and the authorsÕ data provide the variation of the gap function. A fourfold symmetric electron density pattern is observed around the $\Gamma$ and M points. This property could be intrinsic, such as in a weak-SDW \cite{Ran} mode, or arise from the superposition of twin domains expected to form under the structural distortion and SDW transitions. The latter scenario was originally proposed in Ref. \cite{Hsieh}.

The occurrence of Dirac fermions in the system implies that the SDW magnetic state is always metallic. The experimental observations, however, cannot distinguish between a weak-coupling SDW and an antiferromagnetic picture in which some more localized bands are also present. The discovery of the Dirac cone shows that theoretical models that neglect orbital symmetries of iron-based superconductors may not explain the important aspects of the physics. More experiments are needed to address several puzzles. The 122 materials are known to have rather strong $z$-axis dispersion and the evolution of the Dirac cone with the $z$-axis momentum needs to be studied. In several materials, the SDW state coexists with the superconducting state. One possibility for the superconducting order parameter is that of the sign changing s$^\pm$-wave type, which is nodeless. If the superconductor is indeed of the nodeless s$^\pm$ type, does the gapless Dirac fermion still survive in the phase where SDW and superconductivity co-exist? If it does not, then could it be that the superconducting gap opening of the Dirac cone gives rise to a superconductor with topological properties \cite{Wray}? The observation of a Dirac fermion in iron-based materials introduces a new system with protected cones, after cuprate superconductors \cite{Damascelli}, graphene \cite{Geim}, and topological insulators \cite{hasan}. It is likely to inspire new research and unravel the connection between nodal SDW magnetism and the nature of superconductivity.

Iron-based superconductors are the latest example of the Dirac physics that has recently energized condensed matter physics. In cuprate superconductors, Dirac nodes (points where the two cones meet) appear in the structure of the superconducting order parameter, caused by the electron-electron interactions in the parent magnetic Mott insulator. In graphene, degenerate Dirac fermions appear in the noninteracting band structure due to the ${C_3}$ symmetry of the lattice, along with inversion and time-reversal symmetry. In topological insulators, spin-polarized Dirac fermions appear in the strong spin-orbit coupled band structure at the edge of the insulating bulk and enjoy extended protection due to time-reversal symmetry \cite{hasan}. In iron-based superconductors, Dirac fermions appear in the magnetic SDW state band structure \cite{Richard} as a result of the point-group symmetry of the orbitals \cite{Hsieh} making up the Fermi surfaces involved in the nesting process. Even though they enjoy different degrees of protection against disorder and interactions, the Dirac fermions present in all these materials are fascinating examples of how different systems can lead to similar profound low-energy electron behavior.


\begin{thebibliography}{21}


\bibitem{Bednorz} J. G. Bednorz and K. A. M\"uller, {Z. Phys. B: Condens. Matter} {\bf 64}, 189 (1986).
\bibitem{Kamihara} Y. Kamihara, T. Watanabe, M. Hirano, and H. Hosono, {J. Am. Chem. Soc.} {\bf 130}, 3296 (2008).
\bibitem{Ren} Z. A. Ren {\it et al.}, {Chin. Phys. Lett.} {\bf 25}, 2215 (2008).
\bibitem{Cruz}C. de la Cruz {\it et al.}, {Nature} {\bf 453}, 899 (2008).
\bibitem{Richard}P. Richard, K. Nakayama, T. Sato, M. Neupane, Y-M. Xu, J. H. Bowen, G. F. Chen, J. L. Luo, N. L. Wang, X. Dai, Z. Fang, H. Ding, and T. Takahashi, {Phys. Rev. Lett.} {\bf 104}, 137001 (2010).
\bibitem{Ran}Y. Ran {\it et al.}, {Phys. Rev. B} {\bf 79}, 014505 (2009).
\bibitem{Hsieh} D. Hsieh {\it et al.}, arXiv:0812.2289v1.
\bibitem{Hsieh1}D. Hsieh {\it et al.}, {Nature} {\bf 452}, 970 (2008).
\bibitem{Xia} Y. Xia {\it et al.}, {Nature Phys.} {\bf 5}, 398 (2009).
\bibitem{hasan}M. Z. Hasan and C. L. Kane, Rev. Mod. Phys. {\bf 82}, 3045-3067 (2010).
\bibitem{Bernevig}B. A. Bernevig, T. L. Hughes, and S. C. Zhang, {Science} {\bf 314}, 1757 (2006).
\bibitem{Geim} A.V. Geim and K. S. Novoselov, {Nature Mater}. {\bf 6}, 183 (2007).
\bibitem{Damascelli} A. Damascelli, Z. Hussain, and Z.-X. Shen, {Rev. Mod. Phys.} {\bf 75}, 473 (2003); J. C. Campuzano, M. Norman, and M. Randeria in Superconductivity: Physics of Conventional and Unconventional Superconductors,Vol. 1, edited by K. H. Bennemann and J. B. Ketterson (Springer, Berlin, 2008)[Amazon][WorldCat].
\bibitem{Mazin}I. I. Mazin {\it et al.}, {Phys. Rev. Lett.} {\bf 101}, 057003 (2008).
\bibitem{Fang} C. Fang {\it et al.}, {Phys. Rev. B} {\bf 77}, 224509 (2008).
\bibitem{Si} Q. Si and E. Abrahams, {Phys. Rev. Lett.} {\bf 101}, 076401 (2008).
\bibitem{Singh} D. J. Singh and M. H. Du, {Phys. Rev. Lett.} {\bf 100}, 237003 (2008).
\bibitem{Kuroki}K. Kuroki {\it et al.}, {Phys. Rev. Lett.} {\bf 101}, 087004 (2008).
\bibitem{Graser}S. Graser {\it et al.}, {New J. Phys.} {\bf 11}, 025016 (2009).
\bibitem{Wray} L. Wray {\it et al.}, {Nature Phys.} {\bf 6}, 855 (2010).



\end{thebibliography}
\end{document}